# STRUCTURE, MAGNETIC SUSCEPTIBILITY AND SPECIFIC HEAT OF THE SPIN-ORBITAL-LIQUID CANDIDATE $FeSc_2S_4$: INFLUENCE OF Fe OFF-STOICHIOMETRY

V. Tsurkan,[1,2] L. Prodan,[1] V. Felea,[1] I. Filippova,[1] V. Kravtsov,[1] A. Günther,[2] S. Widmann,[2] H.-A. Krug von Nidda,[2] J. Deisenhofer,[2] and A. Loidl[2]

[1]*Institute of Applied Physics, Academy of Sciences of Moldova, MD-2028 Chisinau, Republic of Moldova*
[2]*Experimental Physics V, Center for Electronic Correlations and Magnetism, University of Augsburg, 86135 Augsburg, Germany*

We report structural, susceptibility and specific heat studies of stoichiometric and off-stoichiometric poly- and single crystals of the *A*-site spinel compound $FeSc_2S_4$. In stoichiometric samples no long-range magnetic order is found down to 1.8 K. The magnetic susceptibility of these samples is field independent in the temperature range 10 - 400 K and does not show irreversible effects at low temperatures. In contrast, the magnetic susceptibility of samples with iron excess shows substantial field dependence at high temperatures and manifests a pronounced magnetic irreversibility at low temperatures with a difference between ZFC and FC susceptibilities and a maximum at 10 K reminiscent of a magnetic transition. Single crystal x-ray diffraction of the stoichiometric samples revealed a single phase spinel structure without site inversion. In single crystalline samples with Fe excess besides the main spinel phase a second ordered single-crystal phase was detected with the diffraction pattern of a vacancy-ordered superstructure of iron sulfide, close to the 5C polytype $Fe_9S_{10}$. Specific heat studies reveal a broad anomaly, which evolves below 20 K in both stoichiometric and off-stoichiometric crystals. We show that the low-temperature specific heat can be well described by considering the low-lying spin-orbital electronic levels of $Fe^{2+}$ ions. Our results demonstrate significant influence of excess Fe ions on intrinsic magnetic behavior of $FeSc_2S_4$ and provide support for the spin-orbital liquid scenario proposed in earlier studies for the stoichiometric compound.

## I. Introduction

In search for quantum states of matter, frustrated magnetic $AB_2X_4$ spinels have been intensively studied during the last decades. Structural complexity of the spinel lattice with two different cationic sites, in combination with inherent geometrical frustration of the pyrochlore *B*-site sublattice [1] and bond frustration of the bipartite diamond-type *A*-site sublattice [2], create a fertile ground for their unconventional and exotic behavior. Among the ternary spinels, $FeSc_2S_4$ holds a special place, being so far a prime candidate to manifest spin-orbital liquid (SOL) behavior, an exotic quantum state with entangled spin and orbital degrees of freedom. First magnetic studies of $FeSc_2S_4$ did not find magnetic ordering down to 4.2 K [3]. Further magnetic and specific-heat studies - extended down to 50 mK - did not reveal any long-range magnetic order, too [4]. At the same time, the magnetic susceptibility of $FeSc_2S_4$ was found to follow a Curie-Weiss (CW) behavior in a very broad temperature range from 400 K down to 20 K with a high value of the CW temperature $\theta_{CW} \approx$ -45 K [3,4], indicating the dominance of antiferromagnetic exchange. The frustration parameter, i.e., the ratio of the CW temperature to the ordering temperature for this compound is above 900, being one of the highest values reported so far for frustrated magnets. The lack of long-range magnetic order together with the absence of Jahn-Teller (JT) ordering at low temperatures anticipated for the orbitally degenerate tetrahedral $Fe^{2+}$ ions allowed the authors of Ref. 4 to assign the ground state of this compound to a spin-orbital liquid. Neutron scattering studies of $FeSc_2S_4$ in zero field [5] confirmed the absence of long-range magnetic and of static JT order, revealing strongly dispersive spin excitations with an excitation gap of the order of 0.2 meV. The absence of long-range magnetic order in $FeSc_2S_4$ was also proved by a muon spin-rotation [6]. Nuclear magnetic resonance studies of $FeSc_2S_4$ [7,8] have noted deviations from the CW behavior at low temperatures and a levelling-off of the local magnetic susceptibility below 10 K, reminiscent of strongly frustrated spin systems. Terahertz and far-infrared investigations in zero field of the low-lying excitations in $FeSc_2S_4$ provided clear experimental evidence for the existence of a spin-orbiton excitation related to entangled spin and orbital degrees of freedom [9]. Subsequent terahertz spectroscopy studies under applied magnetic fields evidenced a threefold splitting of this excitation which was attributed to a singlet-triplet transition from the SOL ground state [10]. Recent inelastic neutron-scattering measurements of $FeSc_2S_4$ under applied magnetic fields revealed the shift of the low-energy spectral weight to high energies upon increasing field which was explained by the entangled spin-orbital character of the magnetic states [11]. Theoretical studies of the SOL states [12,13] predicted $FeSc_2S_4$ to lie near a quantum-critical point that separates the spin-orbital singlet and the spin and orbitally ordered state. In contrast to the above mentioned experimental studies, recent neutron-scattering experiments [14] found a long-range antiferromagnetic order with very small magnetic moment below 10 K and an orbitally ordered state below 80 K, thus questioning the magnetic behavior of $FeSc_2S_4$ reported earlier in Refs. 4-10. The composition of samples studied in Ref. 14 was refined with 6% of Fe on the octahedral sites. These samples showed a significant temperature-independent susceptibility which was attributed to ferromagnetic-like impurities, presumably due to iron sulfide phase as reported by subsequent study on single crystals grown by floating zone method [15]. In earlier measurements [4,7] however, no noticeable temperature-independent susceptibility in $FeSc_2S_4$ was found. At the same time, in previous [4,7] and recent [11,14,15] studies of $FeSc_2S_4$ a level below 3% for site intermixing of Fe and Sc ions between the tetrahedral and octahedral sites of the spinel lattice was reported. Within such a low level of inversion it is quite difficult to understand the contrasting magnetic behavior of samples reported in [4,7] and [14,15].

Here we present structural, magnetic and specific-heat studies of $FeSc_2S_4$ performed on poly- and single crystals with different deviations from the ideal stoichiometry. Our study evidences a detrimental effect of the excess Fe ions on intrinsic magnetic properties of $FeSc_2S_4$. We confirm the absence of the long-range magnetic order in stoichiometric $FeSc_2S_4$ and support previous conclusions concerning its spin-orbital liquid ground state.

## II. Experimental

Polycrystalline $FeSc_2S_4$ was prepared by solid-state synthesis from high-purity elements: Fe (99.99%), Sc (99.9%), and S (99.999%). In an Argon box the starting materials were weighted in a stoichiometric ratio and loaded in quartz ampoules, which were pumped to $10^{-2}$ mbar and then sealed. After the first firing at 900 °C for 150 hours the mixture was reground, pressed into pellets, again closed in ampoules and fired at the same temperature. To reach full reaction, the sintering procedure was repeated several times (up to 7 cycles). After each cycle the samples were checked by SQUID and XRD measurements. To optimize the sulfur content, samples were additionally treated in vacuum and sulfur atmosphere and quenched in ice-water. The off-stoichiometric samples were prepared in a similar way from binary FeS and $Sc_2S_3$ preliminary synthesized from the elements. The low level of inversion reported in all previous studies of $FeSc_2S_4$ indicates that Sc ions hardly penetrate into the tetrahedral $A$-sites. This corroborates with the results of structural investigations of related off-stoichiometric spinel compounds $Fe_{0.85}Sc_{2.1}S_4$ [16] and $Mn_{2.29}Sc_{1.14}S_4$ [17] which did not reveal Sc ions at the tetrahedral $A$-sites but found substantial amount of Fe on the octahedral $B$-sites. Therefore in the preparation of the off-stoichiometric samples we assumed the formation of compositions $Fe_{1+x}Sc_{2-x}S_4$, in which the excess Fe ions will occupy the octahedral $B$-sites.

The single crystals have been grown by chemical transport-reactions using iodine as transport agent. The growth experiments were performed at temperatures between 900 and 950 °C for a period up to 12 weeks for each growth run. Technical details for single crystal growth are given in the Supplemental Material [18].

The chemical composition of the samples was determined with an electron-probe microanalyzer CAMECA SX 50 by wavelength dispersive x-ray spectroscopy (WDS). The errors in determination of the absolute concentrations of the elements were not higher than 1.5% for Fe and 2% for Sc and S.

The single crystal x-ray diffraction was performed at room temperature with a Xcalibur E diffractometer equipped with a CCD area detector and a graphite monochromator utilizing Mo Kα radiation. The analysis of diffraction pattern, the search and refinement of unit cell parameters has been done using CrysAlisPro, Agilent Technologies program suite, version 1.171.37.35 [19]. Final unit cell dimensions were obtained and refined on an entire data set. After collection and integration, the data were corrected for Lorentz and polarization effects and for absorption by multi-scan empirical absorption correction methods. The structures were refined by the full matrix least-squares method based on $F^2$ with anisotropic displacement parameters. All calculations were carried out by the programs SHELXL2014 [20]. The x-ray diffraction of polycrystals was performed with a conventional laboratory x-ray diffractometer (STOE Stadi P) with Cu Kα radiation. The data were analyzed with a standard Rietveld refinement using the FullProfSuite program [21].

Magnetic characterization was performed using a commercial SQUID magnetometer (MPMS-5, Quantum Design) for temperatures between 1.8 K and 700 K and in external magnetic fields up to 5 T. The specific heat was measured with a Physical Properties Measurement System (PPMS, Quantum Design) in the temperature range from 1.8 to 300 K.

### III. Results and discussion

**A. Compositional analysis**

In Tab. 1 the results of the WDS analysis of the composition of several poly- and single crystalline samples are given. Within the accuracy of the analysis, the composition of the polycrystalline samples, prepared from the stoichiometric ratio of the elements, was close to that for the ideal stoichiometry (see also Supplemental Material). Similar results were found for single crystalline samples, although a deficiency of iron, varying between 2 and 5 at.%, was observed in different samples from the same batch. An excess of Fe of ~23 at.% and a respective deficiency of ~23 at.% for Sc compared to ideal stoichiometry was revealed in the off-stoichiometric single crystals. Similar results were found for polycrystalline samples with Fe excess. These data appear to be in a reasonable agreement with the formation of compositions $Fe_{1+x}Sc_{2-x}S_4$ with the excess Fe ions on the octahedral sites. Surprisingly, single crystal x-ray studies of these samples (presented below) contradict this intuitive assumption.

Table 1. Composition of $Fe_{1+x}Sc_{2-x}S_4$ samples determined by WDS analysis.

| Sample | Fe (Wt.%) | Sc (Wt.%) | S (Wt.%) | Sum (Wt.%) | Composition |
|---|---|---|---|---|---|
| Ideal stoichiometry $x=0$ | 20.38 | 32.81 | 46.81 | 100.00 | $FeSc_2S_4$ |
| ATF526 $x=0$, poly | 20.65(47)* | 33.27(55) | 46.90(50) | 100.82(83) | $Fe_{1.006(19)}Sc_{2.000(33)}S_{3.977(29)}$ |
| ATR236M1 $x=-0.02$, mono | 20.25(15) | 32.48(34) | 46.18(17) | 98.91(36) | $Fe_{1.003(10)}Sc_{2.000(14)}S_{3.991(10)}$ |
| ATR236M5 $x=0$, mono | 20.24(22) | 32.42(43) | 46.03(31) | 98.69(62) | $Fe_{0.992(12)}Sc_{2.000(21)}S_{3.928(19)}$ |
| ATR215M2 $x=-0.02$, mono | 20.04(19) | 33.22(21) | 46.55(26) | 99.81(56) | $Fe_{0.980(8)}Sc_{2.020(13)}S_{3.974(11)}$ |
| ATR215M3 $x=-0.05$, mono | 19.50(13) | 32.98(28) | 46.71(20) | 99.19(43) | $Fe_{0.952(6)}Sc_{2.000(14)}S_{3.965(16)}$ |
| ATF522 $x=0.06$, poly | 22.06(16) | 31.88(12) | 45.75(29) | 99.69(38) | $Fe_{1.073(7)}Sc_{1.927(10)}S_{3.881(12)}$<br>$0.972FeSc_2S_4+0.122Fe_{0.9}S$** |
| ATF527 $x=0.12$, poly | 22.37(20) | 31.33(34) | 46.43(45) | 100.13(38) | $Fe_{1.098(9)}Sc_{1.910(19)}S_{3.959(16)}$<br>$0.955FeSc_2S_4+0.158Fe_{0.9}S$** |
| ATF540 $x=0.24$, poly | 24.26(31) | 29.44(34) | 45.71(44) | 99.41(67) | $Fe_{1.190(15)}Sc_{1.794(20)}S_{3.906(38)}$<br>$0.897FeSc_2S_4+0.326Fe_{0.9}S$** |
| ATR268M1 $x=0.22$, mono | 24.83(30) | 28.94(17) | 46.40(16) | 100.17(31) | $Fe_{1.218(13)}Sc_{1.764(11)}S_{3.965(16)}$<br>$0.882FeSc_2S_4+0.362Fe_{0.9}S$** |
| ATR268M3 $x=0.23$, mono | 25.04(23) | 29.08(20) | 46.45(33) | 100.57(21) | $Fe_{1.229(12)}Sc_{1.773(14)}S_{3.969(17)}$<br>$0.886FeSc_2S_4+0.380Fe_{0.9}S$** |

*Standard deviations are given in brackets. **Calculated composition for two coexisting phases.

**B. x-ray diffraction**

In Fig. 1 (a) and (b) the x-ray diffraction patterns are presented for single crystals with ideal stoichiometry composition ($x=0$) and off-stoichiometric composition (with iron excess $x=0.23$), respectively. The diffraction pattern for the stoichiometric sample corresponds to a single phase consistent with the spinel structure of $Fd\bar{3}m$ symmetry (# 227). In contrast, the diffraction pattern for the sample with off-stoichiometric composition $x=0.23$ reveals two clearly distinct phases. The analysis of diffraction pattern shows that 1558 of total 1721 observed diffraction reflections correspond to a spinel phase (about 90.7%), Fig. 2(a), and 159 (about 9.3%) to a minority phase, the reflections of which also form an ordered pattern shown in Fig. 2(b). The second phase has been indexed in a cubic system with a rather high value of the lattice constant of 14.82 Å. In fact, for the essentially longer exposure time of data collection the reflections of smaller intensity became visible and regular patterns with even higher lattice constant $a \approx 29.74$ Å were observed. Such a regular pattern of the minority phase reminds of the ordered-vacancy structure characteristics of many binary iron sulfides and selenides. Our magnetic study of samples with Fe excess presented in the following section reveals a ferromagnetic-like contribution of the impurity phase at high temperatures which we primarily associate with a ferrimagnetic NiAs-type $Fe_7S_8$ phase by analogy to Ref. 15. Nonstoichiometric $Fe_{1-\delta}S$ is known to manifest a variety of different superstuctures depending on the preparation conditions and on deviation from the stoichiometry [22,23]. We noticed also that the lattice constant of the second phase in our off-stoichiometric sample is close to that of the 5C polytype of pyrrhotite $Fe_9S_{10}$ [24].

Details of the structural refinement together with the crystal parameters for different single crystalline samples are given in Tables 2 and 3. The crystal structure of the stoichiometric samples corresponds to a normal spinel with Fe ions exclusively at 8$a$ sites and Sc ions at 16$d$ sites. The refinement of site occupation factors for Sc and Fe positions reveals deviations from the expected values (0.08333 and 0.04167, respectively) of less than two and about three *e.s.d.* (estimated standard deviations). Thus, within the accuracy of refinement no site inversion was detected for these samples. Similar results were obtained for samples with Fe deficiency and Fe excess, where intensity of diffraction reflections has been obtained by integration over positions of the peaks corresponding to spinel phase. The deviation of site occupation factors for Sc and Fe positions from the expected values for the spinel phase do not exceed two *e.s.d.* We also noticed enhanced thermal displacement parameters for ions in samples with Fe excess, compared to that for the stoichiometric samples. This additional feature in samples with Fe excess can be related to the influence of the second (impurity) phase.

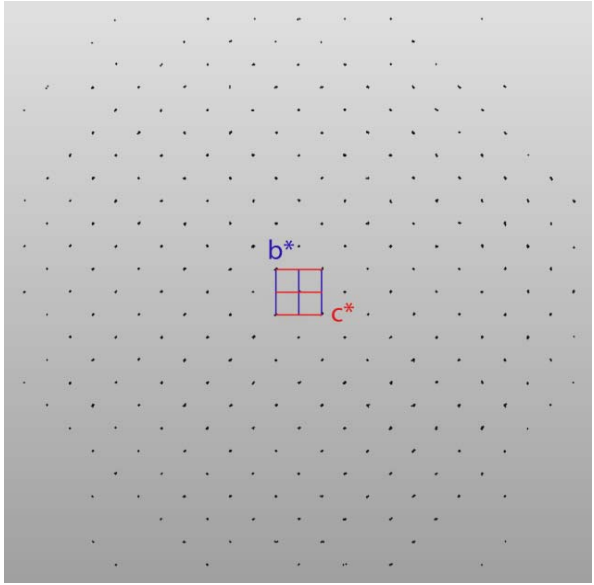

Fig. 1(a) (color online). Image of experimental x-ray diffraction pattern for a stoichiometric single crystal ($x=0$).

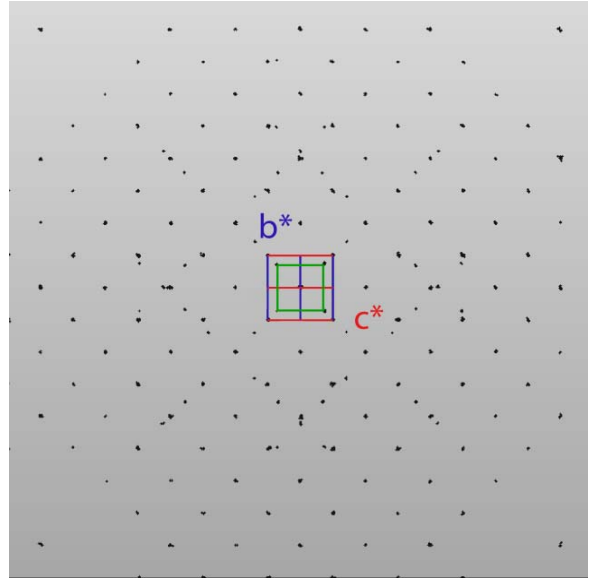

Fig. 1(b) (color online). Image of experimental x-ray diffraction pattern for an off-stoichiometric single crystal with Fe excess ($x=0.23$). Green color corresponds to $b^*$ and $c^*$ axes of the unit cell of the minority phase.

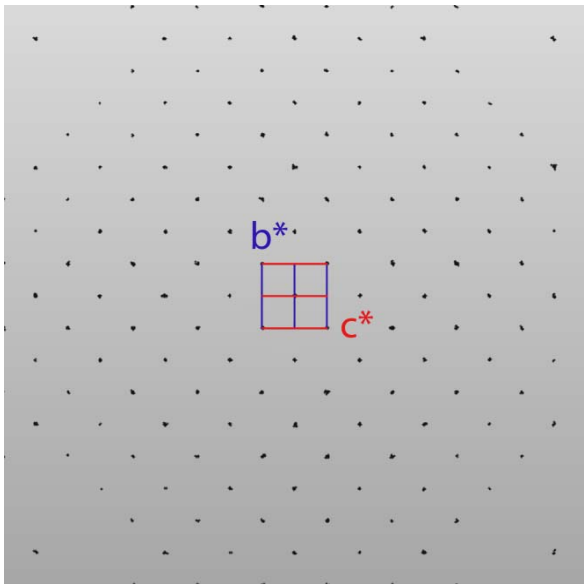

Fig. 2(a) (color online). Image of diffraction pattern for the main phase of the off-stoichiometric single crystal with Fe excess ($x=0.23$).

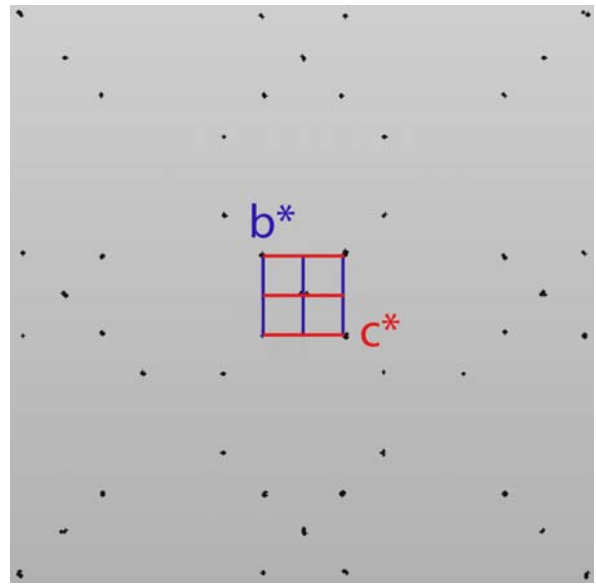

Fig. 2(b) (color online). Image of diffraction pattern for the second phase of the off-stoichiometric single crystal with Fe excess ($x=0.23$).

Table 2. Crystallographic data and details on the structure refinement for the single crystalline samples.

| Sample | ATR236 | ATR268 M1 | ATR215 |
|---|---|---|---|
| Nominal composition | $FeSc_2S_4$ | $Fe_{1.23}Sc_{1.78}S_4$ | $FeSc_2S_4$ |
| Refined composition | $FeSc_2S_4$ | $FeSc_2S_4$ (main phase) | $Fe_{0.98}Sc_2S_4$ |
| Space group | $Fd\bar{3}m$ | $Fd\bar{3}m$ | $Fd\bar{3}m$ |
| $a$ (Å) | 10.5191(2) | 10.5122(2) | 10.5122(2) |
| Sulfur fractional coordinate, $x_0$ | 0.2554(1) | 0.2555(1) | 0.2555(1) |
| Volume (Å$^3$) | 1163.9(1) | 1161.7(1) | 1161.7(1) |
| Crystal size (mm) | 0.3 x 0.2 x 0.15 | 0.33 x 0.23 x 0.2 | 0.35 x 0.3 x 0.2 |
| Reflections collected / unique | 4727 / 108 $R_{int} = 0.0511$ | 5072 / 105 $R_{int} = 0.0331$ | 4671 / 105 $R_{int} = 0.047$ |
| θ range of data collection (º) | 3.355 - 29.996 | 3.357 - 29.790 | 3.357 - 29.791 |
| $\rho_{calc}$ / g cm$^{-3}$ | 3.127 | 3.133 | 3.118 |
| μ / mm$^{-1}$ | 6.006 | 6.018 | 5.960 |
| Transmission min/max | 0.396/1.000 | 0.738/1.0000 | 0.490/1.000 |
| GooF | 1.000 | 1.001 | 1.001 |
| Final $R_1$ [$I>2\sigma(I)$], $wR_2$ | 0.0173, 0.0649 | 0.0184, 0.0754 | 0.0182, 0.0518 |
| $\Delta\rho_{max}$, $\Delta\rho_{min}$ (Å$^{-3}$) | 0.386, -0.892 | 0.647, -0.744 | 0.345, -0.575 |

Table 3. Anisotropic displacement parameters and equivalent isotropic displacement parameters (Å$^2$*10$^3$) for samples with different composition. The displacement factor exponent takes the form: $-2\pi^2[h^2 a^{*2} U_{11} +... + 2hka^*b^*U_{12}]$.

| ATR236 ($x$=0) | | | | | | | |
|---|---|---|---|---|---|---|---|
| Atom | $U_{11}$ | $U_{22}$ | $U_{33}$ | $U_{23}$ | $U_{13}$ | $U_{12}$ | $U_{eq}$ |
| Fe | 12(1) | 12(1) | 12(1) | 0 | 0 | 0 | 12(1) |
| Sc | 10(1) | 10(1) | 10(1) | -1(1) | -1(1) | -1(1) | 10(1) |
| S | 10(1) | 10(1) | 10(1) | -1(1) | -1(1) | -1(1) | 10(1) |
| ATR215 ($x$=-0.02) | | | | | | | |
| Fe | 11(1) | 11(1) | 11(1) | 0 | 0 | 0 | 11(1) |
| Sc | 10(1) | 10(1) | 10(1) | -1(1) | -1(1) | -1(1) | 10(1) |
| S | 10(1) | 10(1) | 10(1) | -1(1) | -1(1) | -1(1) | 10(1) |
| ATR268 ($x$=0.23) | | | | | | | |
| Fe | 18(1) | 18(1) | 18(1) | 0 | 0 | 0 | 18(1) |
| Sc | 16(1) | 16(1) | 16(1) | -1(1) | -1(1) | -1(1) | 16(1) |
| S | 16(1) | 16(1) | 16(1) | 0(1) | 0(1) | 0(1) | 16(1) |

Fig. 3 presents the x-ray powder diffraction patterns for polycrystalline samples $Fe_{1+x}Sc_{2-x}S_4$ with different $x$. All spectra surprisingly show only reflections characteristic of the spinel structure without noticeable impurity peaks. Only in the sample with the highest iron excess $x=0.24$ a small peak centered at $2\theta=43.5^o$ (marked by an asterisk) becomes pronounced over the background. Its angular position correlates well with the maximal peak of the x-ray diffraction spectra for $Fe_7S_8$ shown at the bottom of Fig. 4. On increasing $x$, broadening of the reflections (shown in the inset of Fig. 3) and progressive growth of the background level take place.

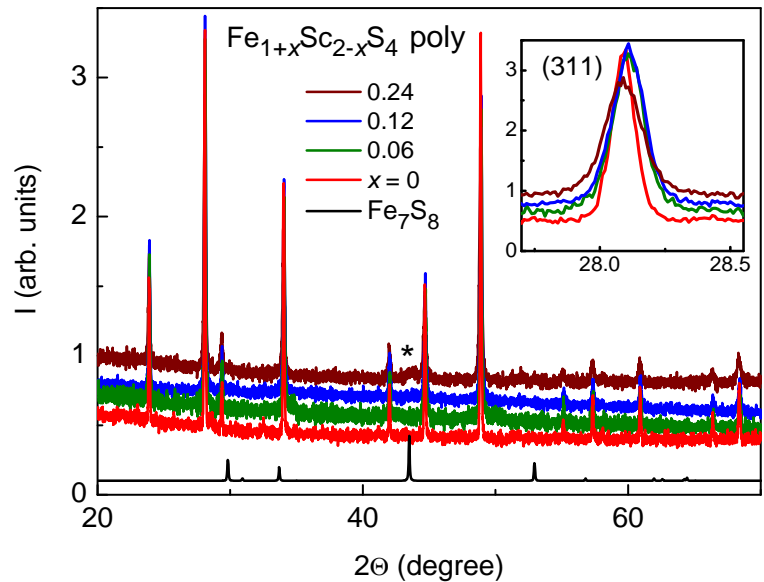

Fig. 3 (color online). x-ray diffraction patterns for polycrystalline samples $Fe_{1+x}Sc_{2-x}S_4$ with different $x$ and of $Fe_7S_8$ (black bottom line). The asterisk marks the impurity peak in sample with $x=0.24$. The inset shows the (311) reflections on enlarged scale demonstrating broadening of the reflections and increasing background in samples with increasing $x$.

In Fig. 4 the x-ray diffraction patterns for crushed single crystalline samples with different $x$ are shown. Like in polycrystals, stoichiometric single crystalline samples and samples with slight iron deficiency ($x=-0.02$) show lines much narrower compared to samples with iron excess $x=0.23$. In the later off-stoichiometric sample also only a single impurity peak of small intensity was detected like in polycrystalline sample with $x=0.24$ although the single crystal x-ray diffraction documented a significant amount (of ~9%) of the second phase. The absence of visible impurity reflections in powder diffraction spectra of the off-stoichiometric samples containing substantial amount of impurity phase (revealed by susceptibility and single-crystal x-ray studies) can explain the apparently contradictory results reported in Refs.14 and 15 and our results concerning the stoichiometry and phase content of the studied samples.

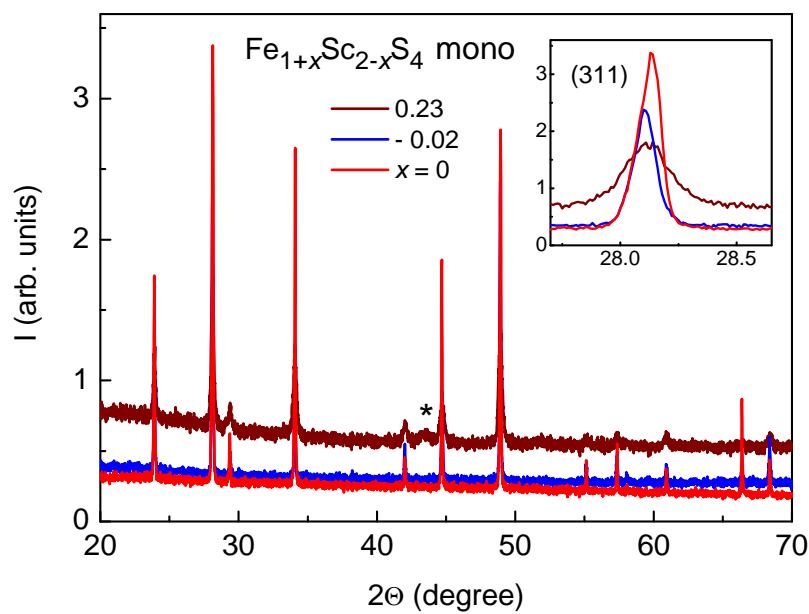

Fig. 4 (color online). x-ray diffraction patterns for crushed single crystalline samples $Fe_{1+x}Sc_{2-x}S_4$ with different *x*. The asterisk marks the impurity peak in sample with *x*=0.23. The inset presents the (311) reflections on enlarged scale showing broadening of the reflection and increased background in sample with *x*=0.23 compared to samples with *x*=0.

In Fig. 5 (a-d) the x-ray diffraction patterns for stoichiometric sample and off-stoichiometric samples with iron excess are shown together with the Rietveld refinement and difference patterns. Within the refinement, the following parameters have been fitted: scale factor, zero point shift, resolution parameters, lattice constant, sulfur positional parameter $x_0$, the site occupations of cations, thermal parameters, asymmetry and preferred orientation. Several structural models were tested for refinement: a) normal spinel with Fe ions only at 8*a* sites and Sc ions at 16*d* sites; b) inversion model with intermixing of Fe and Sc between 8*a* and 16*d* sites; c) Fe distributed between 8*a*, 16*d* and 16*c* sites, and Sc between 16*d* and 16*c* sites proposed previously in Refs. 16 and 17; d) two phase model with a main spinel phase of stoichiometric composition and an impurity phase of iron sulfide.

Applying the inversion model for refinement of the diffraction pattern for polycrystalline samples and powdered single crystalline samples with *x*=0 an intermixing of cations between the *A*- and *B*-sites up to a level to 3% was obtained. Similar low level of inversion of 2.8(8)% in one of the stoichiometric polycrystalline samples (ATF 526) was established by recent high-resolution x-ray and neutron-powder diffraction studies performed in Ref. 11. However, taking into consideration the results of the single-crystal diffraction of the stoichiometric samples, we think that the calculated level of inversion characterizes rather the accuracy of refinement of the powder pattern than the real inversion which appears to be negligible. Therefore the diffraction pattern for the powdered samples with $x=0$ was finally refined assuming full occupation of Fe and Sc respectively, at the tetrahedral $8a$ and octahedral $16d$ sites as follows from the single-crystal diffraction.

The refinement for samples with Fe excess did not reveal Fe ions at $16c$ sites of the spinel lattice. Within the inversion model, a level up to 4% of Fe at $16d$ sites was obtained. No respective amount of Sc ions at the $A$-sites was found. The refinement for the polycrystalline samples with Fe excess within the two-phase model was done assuming the presence of iron sulfide as the second phase, with a hexagonal $P6_3/mmc$ or monoclinic $P12/c$ symmetry. Within the two-phase model we got only slightly better $R$ factors for the main $FeSc_2S_4$ phase compared to other tested models. However, the Bragg-$R$ factor and $R_f$ factor for the second phase were rather high and due to poor quality of the refinement it was not possible to draw definite conclusions concerning the second phase.

The obtained lattice parameters, fractional coordinates of sulfur, and residuals of refinement for different powdered samples are given in Table 6. The lattice parameters of the spinel phase for samples with different $x$ show only small variation which can be probably attributed to details of their preparation, e.g., minor deviations from the stoichiometry, difference in regimes of thermal treatments, treatment atmosphere, etc. Importantly, rather low scattering of values of the lattice constants for samples with different $x$ indicates high structural stability of the main spinel phase and suggests insignificant level of inversion.

Summarizing the results of the structural analysis, we can fully exclude the inversion for stoichiometric $FeSc_2S_4$ samples. The x-ray diffraction analysis points also to a small inversion degree (probably also absent) in the main spinel phase in samples with Fe excess. The excess Fe ions form a second phase of iron sulfide which intergrows with the main spinel phase.

Based on the results of the structural analysis of the off-stoichiometric samples which provide clear evidence for coexistence of the main stoichiometric spinel phase and the second phase of iron sulfide, we reanalyzed the data of the WDS compositional analysis which are also presented in Table 1. We got a reasonable agreement concerning the amount of the main and the second phase in the off-stoichiometric single crystalline samples with those obtained from the single-crystal diffraction. Within the accuracy of the WDS analysis, the composition of the off-stoichiometric polycrystalline samples can be well described by a two-phase model with decreasing concentration of the main phase as $(1-x/2)\text{FeSc}_2\text{S}_4$ and increasing concentration of the second phase as $(3x/1.8)\text{Fe}_{0.9}\text{S}$ on increasing of Fe excess $x$.

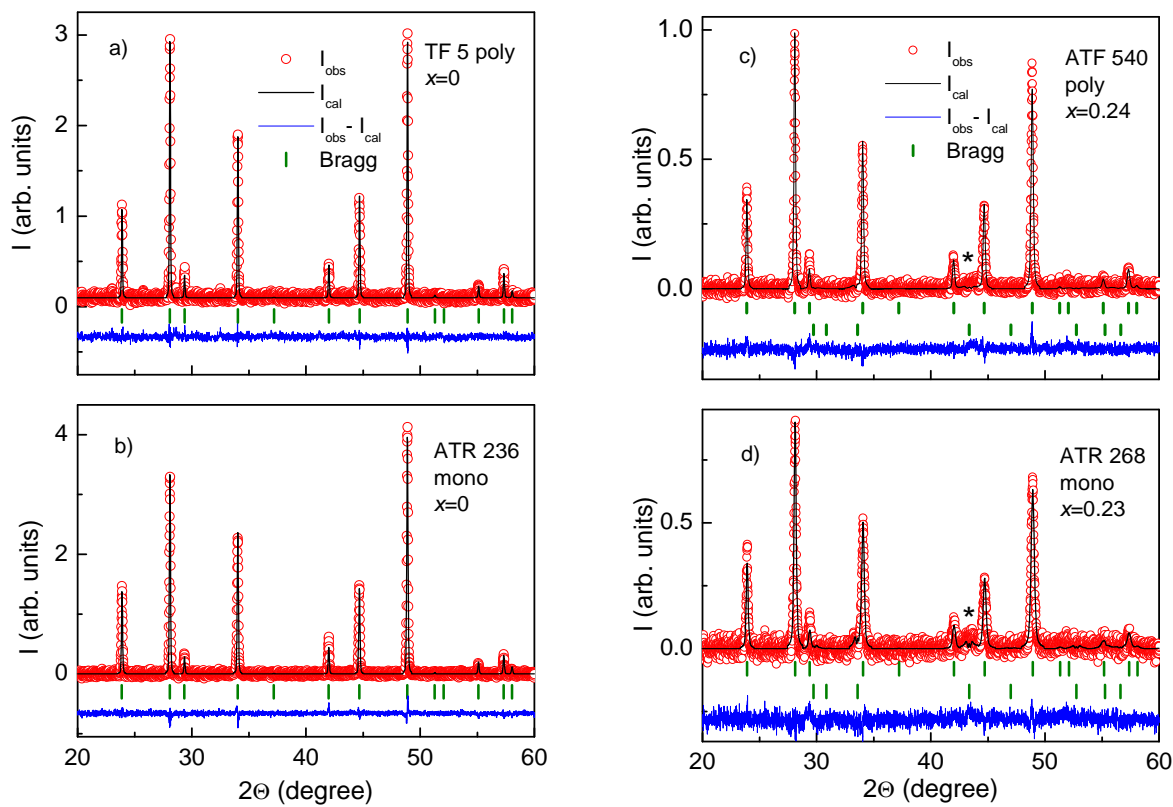

Fig. 5 (color online). x-ray diffraction patterns for samples: a) polycrystalline stoichiometric $\text{FeSc}_2\text{S}_4$; b) crushed single crystalline stoichiometric $\text{FeSc}_2\text{S}_4$; c) polycrystalline off-stoichiometric $\text{Fe}_{1.24}\text{Sc}_{1.76}\text{S}_4$; d) single crystalline off-stoichiometric $\text{Fe}_{1.23}\text{Sc}_{1.78}\text{S}_4$. The measured intensities (red open circles) are compared with the calculated profile (black solid line). Bragg positions of the spinel and of the impurity phases are indicated by vertical (green) bars. The difference pattern is shown by the blue line at the bottom. Background is removed.

Table 6. Lattice constants $a_0$, sulfur positional parameter $x_0$ (in fractional coordinates, f.c.), residuals $R_{wp}$ and Chi$^2$ obtained by Rietveld refinement of the x-ray patterns for different samples.

| Sample | $x_0$ (f.c.) | $a_0$ (Å) | $R_{wp}$ | Chi$^2$ |
|---|---|---|---|---|
| TF 5 ($x$=0) poly | 0.2551(4) | 10.5159(3) | 4.54 | 1.10 |
| ATF 526 ($x$=0) poly | 0.2550(3) | 10.5107(3) | 5.61 | 1.09 |
| ATR 236 ($x$=0) mono | 0.2555(2) | 10.5163(2) | 4.12 | 1.02 |
| ATR215($x$=-0.02) mono | 0.2527(6) | 10.5161(3) | 6.04 | 1.16 |
| ATR 274($x$=-0.02) mono | 0.2541(3) | 10.5147(2) | 4.72 | 1.18 |
| ATF 522 ($x$=0.06) poly | 0.2551(3) | 10.5118(3) | 5.85 | 1.02 |
| ATF 527 ($x$=0.12) poly | 0.2557(2) | 10.5135(2) | 4.42 | 1.06 |
| ATF 540 ($x$=0.24) poly | 0.2548(3) | 10.5194(4) | 3.75 | 1.16 |
| ATR 268 ($x$=0.23) mono | 0.2557(3) | 10.5119(8) | 3.95 | 1.15 |
| ZnSc$_2$S$_4$ poly | 0.2550(2) | 10.4812(1) | 7.61 | 1.23 |

**C. Magnetic properties**

The temperature dependences of the magnetic susceptibility, $\chi$, and of the inverse susceptibility for several stoichiometric FeSc$_2$S$_4$ poly- and single crystals ($x$=0) measured in a field of 1 T are shown in Fig. 6. In agreement with previous studies [3,4], we observed an extended temperature range (from 10 to 400 K) of Curie-Weiss (CW) behavior of the susceptibility for stoichiometric samples with very similar values of $\chi$ for samples from different batches. The susceptibility for these poly- and single crystals amounts ~7.5∗10$^{-3}$ emu/mol at 400 K. For stoichiometric samples no significant temperature independent susceptibility is revealed. Its amount is estimated at a level ~0.3∗10$^{-3}$ emu/mol. These samples contain only a single spinel phase and therefore we can attribute the measured susceptibility to an intrinsic susceptibility of stoichiometric FeSc$_2$S$_4$. Below 6 K, the susceptibility of the stoichiometric polycrystalline sample shows a slight upturn which reminds of a Curie tail due to possible residual magnetic impurities. However, a similar upturn in $\chi$ at low temperatures was observed in our stoichiometric single crystals, where the amount of residual magnetic impurities is anticipated to be at a much lower level than in polycrystals. Note that the susceptibilities of our stoichiometric samples reveal pronounced field dependence only at temperatures below 6 K, as shown in the inset of Fig. 6. The susceptibility levels-off at high fields and at lowest temperatures. At the same time, at temperatures above 6 K, the susceptibility of stoichiometric samples is found to be completely field-independent. In the case of impurities originating from the ferrimagnetic $Fe_{1-\delta}S$ with a Curie temperature of 588 K [25] the susceptibility is expected to be clearly field dependent at high temperatures, where the contribution from the intrinsic susceptibility of the spinel phase $FeSc_2S_4$ is strongly reduced. Such a field-dependent behavior of the susceptibility at high temperatures however was not observed in our stoichiometric samples. Therefore, we conclude that the susceptibility upturn observed in stoichiometric samples at low temperatures and in low fields is not related to $Fe_{1-\delta}S$ impurities. The levelling-off of the local susceptibility at temperatures below 10 K detected in NMR studies of $FeSc_2S_4$ [6,7] correlates well with our susceptibility data for an applied field of 5 T and thus cannot be related to impurities either.

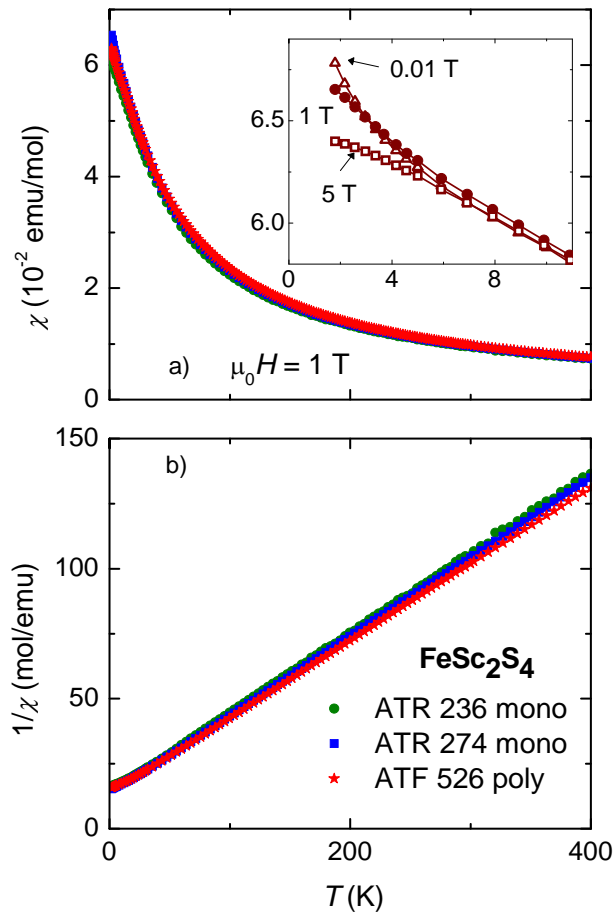

Fig. 6 (color online). Temperature dependences of the magnetic susceptibility (a), and of the inverse susceptibility (b), measured in a magnetic field of 1 T for stoichiometric polycrystalline and single crystalline samples of $FeSc_2S_4$. Inset in (a) shows the susceptibility of the stoichiometric single crystal measured at different applied magnetic fields for the temperature range below 10 K.

In contrast to stoichiometric samples, the susceptibility of samples with iron excess was found to be strongly field-dependent in the temperature range 10 - 400 K. Fig. 7 (a) shows the temperature dependences of the magnetic susceptibilities and Fig. 7 (b) - of the inverse susceptibilities for poly- and single crystals with iron excess measured in a field of 5 T. The application of a higher field was necessary to saturate impurity contribution to the magnetization of these samples which will be further discussed below. The significant difference in values of the susceptibility $\chi$ for fields of 1 and 5 T is demonstrated in Fig. 7 (b). Note also that the temperature dependence of the inverse susceptibility in a field of 1 T is highly non-linear compared to the strictly linear dependence observed in the stoichiometric samples (Fig. 6b). Below 20 K, the susceptibility for polycrystalline samples with Fe excess $x$=0.06 and 0.12 shows a hump at around 10 K and levels-off at lower temperatures at values notably smaller than for stoichiometric samples (see insets in Figs. 6a and 7a). In poly- and single crystalline samples with higher iron excess $x$=0.24, the susceptibility develops a well pronounced maximum at 10 K reminiscent of a magnetic transition. The other important observation is that the values of the susceptibilities of samples with iron excess at high temperatures are significantly higher compared to stoichiometric ones (see Table 7). The difference between the susceptibilities of the off-stoichiometric sample (with $x$=0.24) and stoichiometric samples is demonstrated in Fig. 8. Above 100 K the data for the field of 5 T show an almost temperature independent contribution, and we note that the values of the susceptibilities of the off-stoichiometric samples at high temperatures are similar to those documented in Refs. 14 and 15. There, a temperature independent contribution, $\chi_0$, varying from $3.1 \times 10^{-2}$ emu/mol for polycrystalline samples to $1.77 \times 10^{-1}$ emu/mol for single crystals was reported, values which are 4 to 23 times higher than the intrinsic susceptibilities of our stoichiometric samples at temperatures above 100 K. Taking into consideration a ferromagnetic-like behavior of the susceptibility under magnetic field observed in our poly- and single crystalline samples with iron excess, we attribute it to the second phase revealed by single-crystal diffraction.

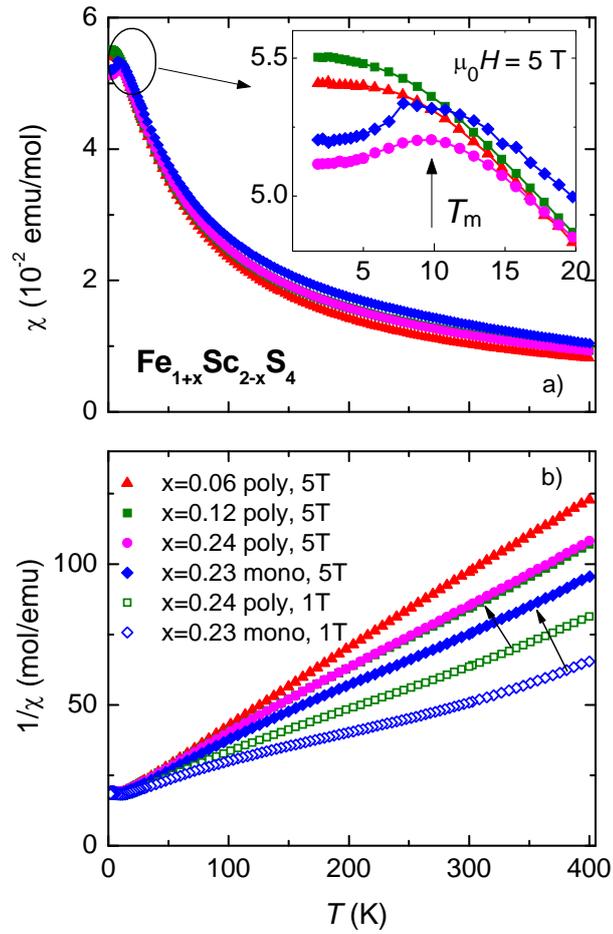

Fig. 7 (color online). Temperature dependences of the magnetic susceptibility (a), and of the inverse susceptibility (b), measured in magnetic fields of 5 T (closed symbols) and 1 T (open symbols) for poly- and single crystalline samples with iron excess. Inset in (a) shows the susceptibility behavior below 20 K on an enlarged scale. The vertical arrow marks the maximum in the susceptibility at $T_m \approx 10$ K. The arrows in (b) show the shift of the $\chi^{-1}(T)$ dependences on increasing field from 1 to 5 T demonstrating strong field dependence of the susceptibility of the samples with Fe excess.

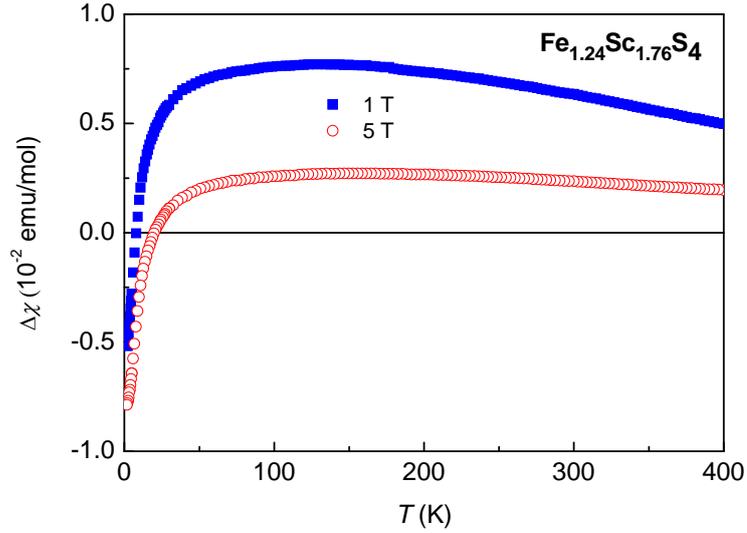

Fig. 8 (color online). Difference in the susceptibilities of the off-stoichiometric sample (with $x=0.24$) and stoichiometric sample ($x=0$) *vs.* temperature in fields of 1 and 5 T demonstrating strong field dependence of the susceptibility and the temperature independent contribution above 100 K for the 5 T data for the off-stoichiometric sample.

To clarify the origin of the second magnetic phase existing in samples with Fe excess, we extended the susceptibility measurements to temperatures up to 700 K. In Fig. 9 we present the temperature dependences of the magnetic susceptibility (a) and inverse susceptibility (b) measured in a magnetic field of 0.1 T for polycrystalline sample with Fe excess $x=0.24$ at temperatures between 300 and 700 K. The data measured on heating document a clear magnetic phase transition at $T_C=580$ K with a thermal hysteresis of ~20 K on cooling cycle. Above $T_C$ the susceptibility of this sample follows a Curie-Weiss law indicating a true paramagnetic state. Note that very similar magnetic behavior was reported in Ref. 25 for ferrimagnetic iron sulfide $Fe_{0.902}S$. These results suggest that the second magnetic phase detected in our off-stoichiometric samples can be associated with the phase $Fe_{1-\delta}S$ with $\delta \approx 0.098$, i.e. a composition close to the 5C pyrrothite $Fe_9S_{10}$.

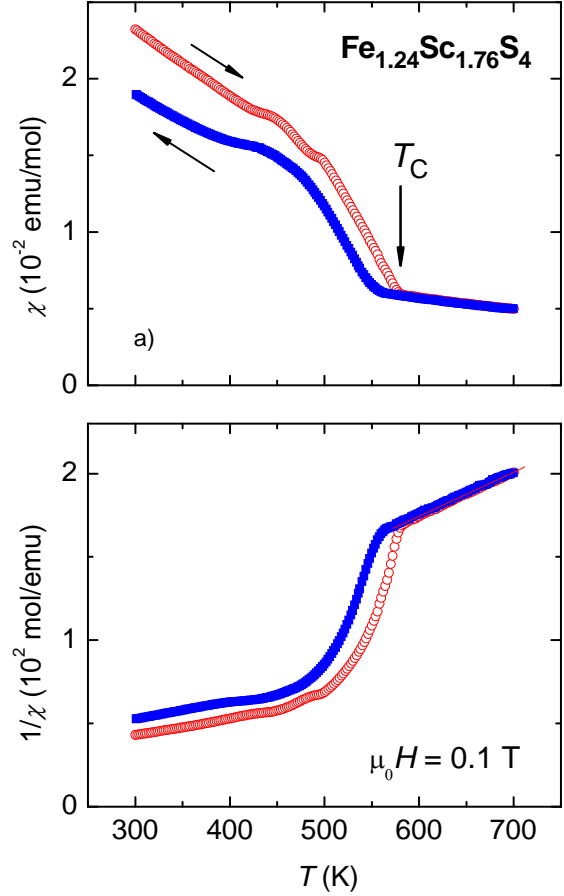

Fig. 9 (color online). Temperature dependences of the magnetic susceptibility (a) and of the inverse susceptibility (b) measured in a magnetic field of 0.1 T for the polycrystalline sample with Fe excess $x=0.24$. Vertical arrow marks the magnetic transition on heating cycle.

Table 7 summarizes the magnetic parameters, specifically the effective magnetic moment, $p_{eff}$, and the Curie-Weiss temperature, $\Theta_{CW}$, determined from the CW fits to the experimental data. For stoichiometric samples, $\Theta_{CW}$ and $p_{eff}$ were calculated for the temperature range 20 - 400 K, where a strict linear temperature dependence of the inverse susceptibility measured in a field of 1 T was observed. The values of these parameters scatter in a narrow range around -45 K and 5.15 $\mu_B$, respectively. The latter is in good agreement with the single-ion value for $Fe^{2+}$ ions with a $g$-factor of 2.07 [26]. The effective magnetic moment and the Curie-Weiss temperature for samples with off-stoichiometric iron were calculated from the data measured in a field of 5 T to reduce the field-dependent effects. The values of $p_{eff}$ and $\Theta_{CW}$ are found to be increased compared to those for the stoichiometric samples. Similar enhanced values of $\Theta_{CW}$ and $p_{eff}$ were reported for crystals studied in Ref. 15. However, in our opinion, the values of $p_{eff}$ and $\Theta_{CW}$ for the off-stoichiometric samples have to be considered with precaution, since our experiments show that the true paramagnetic state in these samples is reached only above 580 K.

Table 7.
Magnetic parameters calculated from the CW fit to the susceptibility data and the value of the susceptibility at 400 K measured in a field of 1 and 5 T.

| Sample label | Type | $p_{eff}$/Fe ($\mu_B$) | $-\Theta_{CW}$ (K) | Susceptibility value at 400 K in 1 T/ 5 T ($10^{-3}$ emu/mol Oe) | Sample from |
|---|---|---|---|---|---|
| N1 | Poly ($x=0$) | 5.12 | 45.1 | 7.3/- | Ref. [4] |
| N2 | Poly ($x=0$) | 5.15 | 45.2 | 7.5/- | Refs. [6,7] |
| N5 | Poly ($x=0$) | 5.18 | 43.2 | 7.7/- | Ref. [9] |
| N7 | Poly ($x=0$) | 5.24 | 45.0 | 7.8/- | Ref. [9] |
| ATF 518 | Poly ($x=0$) | 5.10 | 46.3 | 7.32/7.30 | Present work |
| ATF526 | Poly ($x=0$) | 5.22 | 46.5 | 7.6/- | Ref. [11] |
| ATR 236 | Mono ($x=0$) | 5.12 | 46.7 | 7.3/- | Present work |
| ATR 274 | Mono($x=-0.02$) | 5.14 | 44.5 | 7.64/7.64 | Present work |
| ATF 522 | Poly ($x=0.06$) | 5.27 | 57.5 | 8.26/8.14 | Present work |
| ATF 527 | Poly ($x=0.12$) | 5.61 | 80 | 12.64/9.34 | Present work |
| ATF 540 | Poly ($x=0.24$) | 5.49* | 57* | 12.28/9.24 | Present work |
| ATR 268 | Mono ($x=0.23$) | 5.68 | 72 | 15.28/10.46 | Present work |

*Calculated for the temperature range 600-700 K.

To understand the reason of the different magnetic behavior of samples with stoichiometric and off-stoichiometric composition we focused on their low temperature magnetic properties. Fig. 10 shows the magnetization curves measured at 2 K for samples with different Fe content. Both poly- and single crystalline samples of stoichiometric $FeSc_2S_4$ exhibit a linear increase of the magnetization with field and no hysteretic effects. In contrast, samples with iron excess show a significant hysteresis of magnetization with the appearance of remnant magnetization and coercive field, which both increase with increasing iron excess, as documented in the inset of Fig. 10. The hysteresis loop closes in fields above 3 T, which is comparable with the saturation field of the ferrimagnetic phase of iron sulfide. The temperature dependence of the remnant magnetization, $M_r$, measured on the sample with the highest $x=0.24$ reveals a sharp decrease of $M_r$ at 10 K and an hump at around 20 K followed by a continuous decrease at high temperatures (see Supplemental Material). Even at 400 K, $M_r$ has a nonzero value vanishing at higher temperatures, probably at $T_C$.

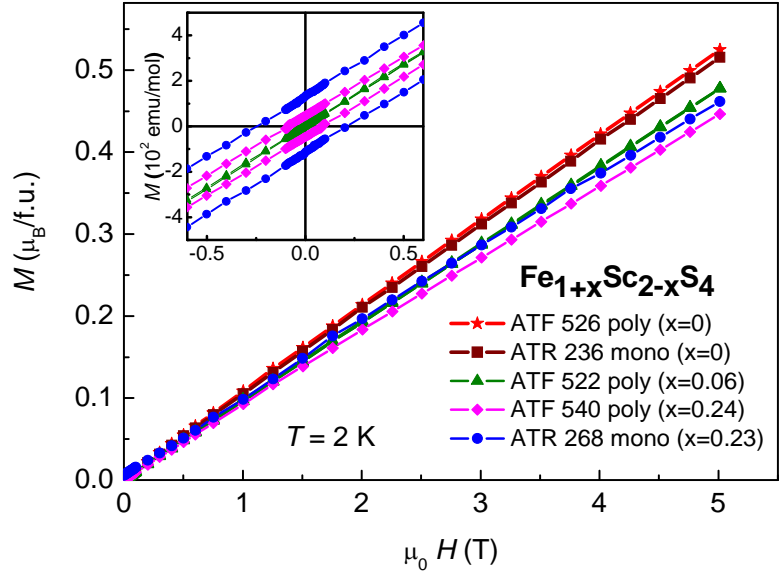

Fig. 10 (color online). Magnetization curves measured at 2 K for samples with different composition. Inset shows hysteresis effects observed in samples with Fe excess.

To have a deeper insight into the low-temperature magnetic properties of samples we performed measurements of the *dc* susceptibility in the low-field range using conventional zero-field cooled (ZFC) and field-cooled (FC) sequences and of the *ac* susceptibility in a frequency range 1-1000 Hz. In Fig. 11 the temperature dependences of the ZFC and FC susceptibilities for several samples measured in a magnetic field of 0.01 T are presented. We observed that the susceptibility for both poly- and single crystalline samples of stoichiometric $FeSc_2S_4$ exhibits a monotonic increase with decreasing temperature. No detectable difference between ZFC and FC data was observed for stoichiometric samples that documents a negligible level of magnetic disorder. Similar behavior was found in samples with iron deficit of 5%. In contrast, strong magnetic irreversibilities were revealed for samples with Fe excess, which were reflected by a pronounced difference between ZFC and FC curves reminiscent of spin-glass-like behavior. The ZFC and FC curves split at approximately 80 K and flatten below 10 K. In the single crystal with the highest concentration of excess iron $x=0.23$, in addition to irreversible behavior, the susceptibility develops a clear maximum at around 10 K. A similar maximum in the susceptibility at 10 K was observed in the polycrystalline sample with iron excess $x=0.24$ (see Supplemental Material). Note that the irreversible behavior of the low-field susceptibilities observed in off-stoichiometric polycrystalline samples is again very similar to that reported for samples in Refs. 14 and 15. Our results thus indicate clear relation of the magnetic irreversibility effects with the second phase formed by excess Fe.

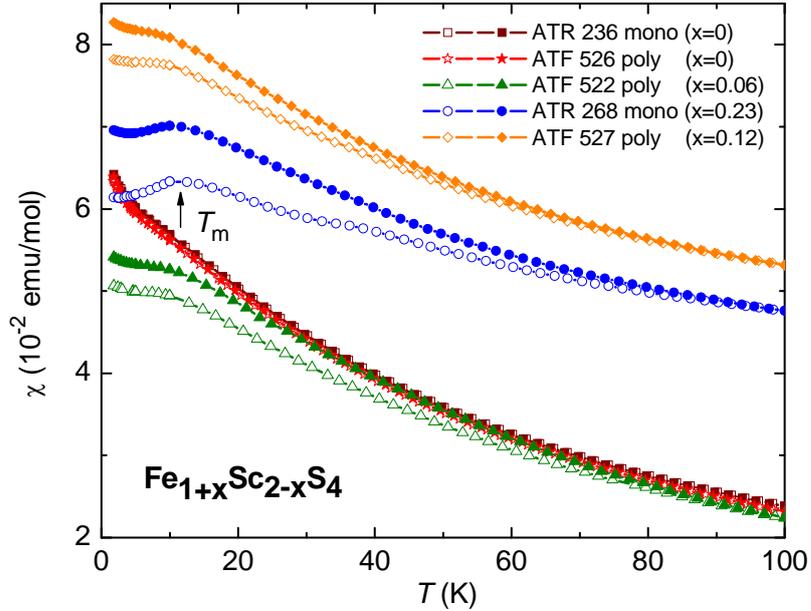

Fig. 11 (color online). Temperature dependences of the *dc* susceptibilities measured in a magnetic field of 0.01 T for poly- and single crystalline samples with different Fe content. Open and closed symbols mark ZFC and FC data, respectively. Vertical arrow marks the maximum at $T_m$.

In Fig. 12 the temperature dependences of the *ac* susceptibilities measured in a frequency range 1-1000 Hz for the polycrystalline sample with stoichiometric iron content $x=0$ and for the single crystalline sample with iron excess $x=0.23$ are presented. In the stoichiometric sample the susceptibility is frequency independent and shows continuous increase on decreasing temperature similar to *dc* data (Fig. 6 a), indicating the absence of any magnetic ordering. In contrast, susceptibility of the sample with iron excess $x=0.23$ exhibits a non-monotonic temperature behavior with a broad maximum centered at 11 K. The susceptibility of this off-stoichiometric sample also is found to be frequency independent. Such a behavior of the susceptibility is in disagreement with that expected for conventional spin glasses or cluster glasses which exhibit pronounced frequency dependence [27]. In fact, the observed behavior of *ac* and *dc* susceptibilities of samples with iron excess below 20 K either suggests the onset of the antiferromagnetic order or anomalous magnetic behavior due to the second phase detected in these samples. The observation of partial antiferromagnetic spin ordering with reduced moment in Ref. 14 seems to favor the first scenario. However, in our opinion, one cannot exclude an anomalous magnetic behavior due to interaction at the interface of the main spinel phase $FeSc_2S_4$ with the secondary phase of $Fe_{1-\delta}S$ revealed in these samples. Very similar magnetic behavior as in off-stoichiometric samples was reported recently for monoclinic pyrrhotite at low temperatures due to interaction at the interface of the two coexisting superstructures 4C and 5C* of $Fe_{6.6}S_8$ resulting from the change of magnetocrystalline anisotropy [28, 29]. Further studies are evidently necessary to discriminate between the two possible scenarios of anomalous magnetic behavior at low temperatures observed in the off-stoichiometric samples.

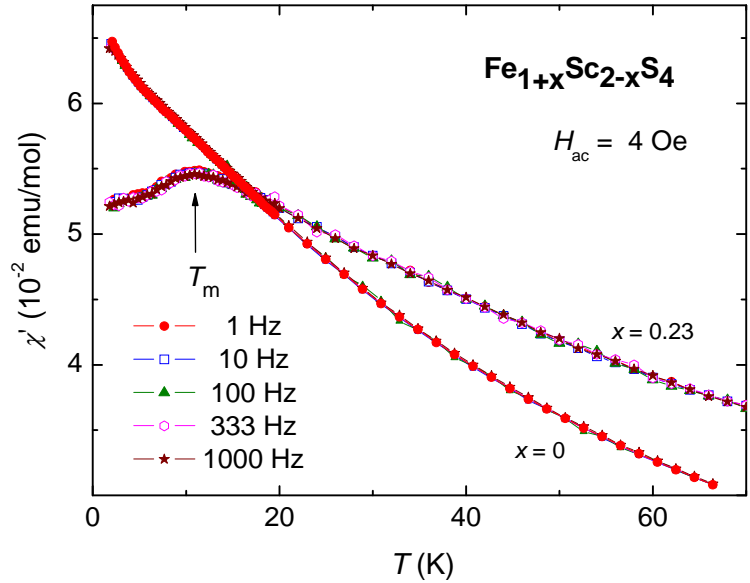

Fig. 12 (color online). Temperature dependences of the *ac* susceptibility measured at different frequencies for polycrystalline sample with stoichiometric iron content *x*=0 and for single-crystalline sample with iron excess *x*=0.23. Vertical arrow marks the maximum at 11 K.

**D. Specific heat**

In Fig. 13 the temperature dependences of the specific heat, *C,* for stoichiometric and off-stoichiometric samples measured in zero field at ambient pressure are shown. The specific heat for all samples reveals a similar continuous decrease on lowering the temperature from 300 to 20 K. No sharp anomaly or jump indicative of long-range magnetic or structural transformations at lower temperatures was found in any sample. Instead, a broad Schottky-like contribution develops at temperatures below 20 K, both for stoichiometric and non-stoichiometric samples. When plotting the specific heat as *C/T vs. T* shown in the inset of Fig. 13 this broad anomaly is centered at 7.37 K, 8.23 K, and 8.40 K, respectively, for stoichiometric ($x=0$), and off-stoichiometric samples with iron excess $x=0.06$ and $x=0.24$. The observed behavior of *C* for stoichiometric samples is in complete agreement with previous report on the specific heat [4]. The behavior of *C/T* for our off-stoichiometric samples is similar to that reported in recent studies in Refs. 14 and 15.

Fig. 14 shows the temperature dependences of the specific heat, in a representation *C/T vs. T* at temperatures up to 200 K. At temperatures below the maximum in *C/T*, the specific heat of the off-stoichiometric samples is lower than that of the stoichiometric samples while above 20 K it shows higher values than for the stoichiometric samples. The dashed line in Fig. 14 presents the lattice contribution to the specific heat of $FeSc_2S_4$, which was calculated from the experimentally measured specific heat of the related spinel $ZnSc_2S_4$. This nonmagnetic compound differs by only 3.5 % in molar mass compared to $FeSc_2S_4$. Using a combined Debye-Einstein model, we were able to describe the specific heat of the non-magnetic $ZnSc_2S_4$ within the accuracy ~0.5% in the temperature range from 40 to 250 K (see Supplemental Material). The calculated phonon contribution is found to be more accurate compared to that estimated from the specific heat data of $CdIn_2S_4$ with a much larger difference in the molar mass (by 72 %) compared to $FeSc_2S_4$. The phonon contribution to the specific heat for the stoichiometric $FeSc_2S_4$ samples for temperatures above 50 K was simulated by one isotropic Debye (*D*) and three isotropic Einstein terms (*E*1, *E*2, and *E*3). The 21 degrees of freedom per formula unit were taken into account by setting the ratio of these contribution to $D : E1 : E2 : E3 = 1 : 1 : 3 : 2$ using the Debye and Einstein temperatures $\theta_D = 183$ K, $\theta_{E1} = 175$ K, $\theta_{E2} = 340$ K, $\theta_{E3} = 497$ K, which are consistent with the frequency ranges where IR active phonons have been observed [9].

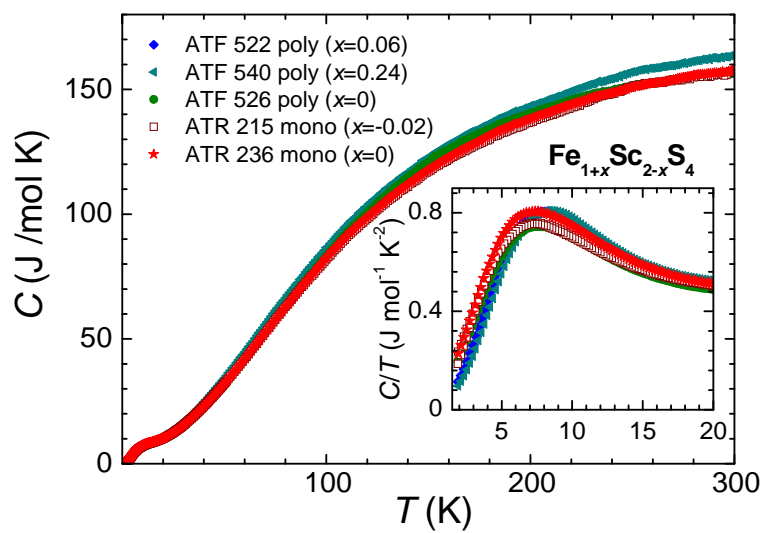

Fig. 13 (color online). Temperature dependences of the specific heat $C$ for stoichiometric and off-stoichiometric samples. Inset presents the low-temperature data in a representation $C/T$ on a enlarged scale, showing the shift of the maximum in $C/T$ to higher temperatures with increasing excess of Fe.

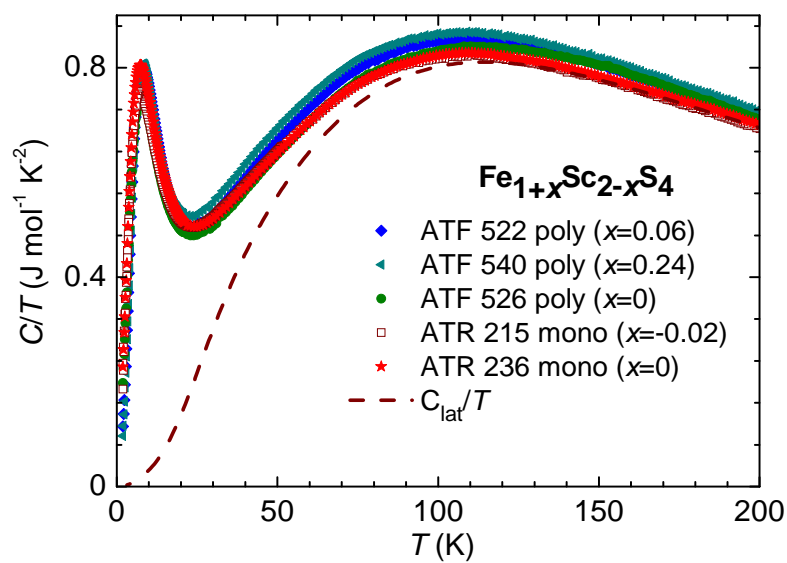

Fig. 14 (color online). Temperature dependences of the specific heat $C$ divided by temperature $T$ for stoichiometric and off-stoichiometric samples. The dashed line marks the lattice contribution.

We analyzed the experimental data by assuming two different contributions to the specific heat, namely, from vibrational degrees of freedom and from the lowest-lying electronic levels of the $Fe^{2+}$ ions. The lattice contribution $C_{lat}$ has been subtracted from the experimentally observed specific heat. The remaining residual specific heat $C_m/T$ for stoichiometric and off-stoichiometric samples and the entropies obtained by integrating $C_m/T$ data are shown in Fig. 15 (a) and (b), respectively. The entropy contained in the residual specific heat for the stoichiometric single crystalline samples saturates close to the value of Rln(10) consistent with that expected for the spin Rln(2S+1) and orbital Rln2 degrees of freedom. The deviations from the value of Rln(10) for the entropy of the non-stoichiometric samples above 100 K can be probably attributed to the second phase.

Finally, in Fig. 15 (a) we show a Schottky-like contribution reflecting the lowest lying electronic levels of the $Fe^{2+}$ ions, where spin and orbital degrees of freedom are coupled giving rise to an entropy of Rln(10). The energy level scheme corresponds to the spin-orbital-singlet ground-state, a first excited triplet, which was reported by THz-spectroscopy and neutron diffraction [9, 11], followed by a doublet, a triplet, and another singlet state. To calculate this electronic contribution we used the expressions

$$C_{low-lying}(T) = N \frac{\partial E}{\partial T} \quad (1)$$

$$E = \frac{1}{Z}\sum_{i=0}^{5} g_i \epsilon_i e^{-\beta \epsilon_i} \quad (2)$$

and the partition function

$$Z = \sum_{i=0}^{5} g_i e^{-\beta \epsilon_i} \quad (3)$$

with the excitation energies $\varepsilon_{0,1,2}$ = 0, 2.5, 8, 11, 13 meV; the degeneracies $g_{0,1,2}$ = 1, 3, 2, 3, 1, and $\beta \equiv 1/k_B T$. Note that only the energy of the first excited triplet has been observed by THz and neutron studies with an energy of about 4.5 meV. By assuming a lower value of 2.5 meV we parametrized the actual triplet dispersion as measured by neutron scattering [5] in a single-ion picture. The resulting curve is in good agreement with the residual specific heat for all samples and accounts well for the entropy at low temperatures.

Summarizing, we conclude that the total specific heat of all samples can be reasonably well explained by phononic contributions and the contribution from the low-lying electronic levels of $Fe^{2+}$ ions, where spin and orbital degrees of freedom are entangled. No ordering phenomena are necessary to explain the experimental specific heat.

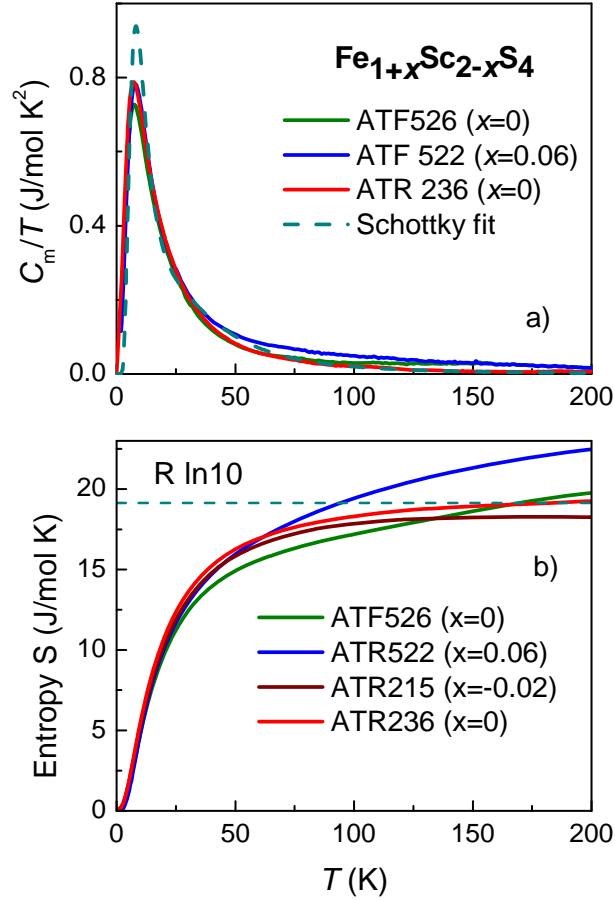

Fig. 15 (color online). Temperature dependences of the magnetic part of the specific heat $C_m$ (a), and of the entropy (b) for stoichiometric and off-stoichiometric samples. Dashed lines marks the entropy expected for the spin Rln(2S+1) and orbital Rln2 degrees of freedom.

### III. Concluding remarks

Our detailed structural, magnetic and specific-heat studies performed on polycrystalline and single crystalline samples of $FeSc_2S_4$ with well controlled composition testify the following important peculiarities of this material:

1. Structural analysis points to an extremely low level of inversion of the spinel phase in both stoichiometric and off-stoichiometric samples. The excess Fe in the off-stoichiometric samples forms a second phase with a regular diffraction pattern corresponding to a vacancy ordered iron sulfide with composition close to the 5C polytype of pyrrhotite ($Fe_9S_{10}$).

2. The susceptibility of the stoichiometric samples ($x=0$) follows a Curie-Weiss law in an extended temperature range from 10 to 400 K and does not reveal any sign of long-range magnetic order down to the lowest measured temperature of 1.8 K. No magnetic irreversibility is detected for these samples, indicating the absence of magnetic disorder.

3. The susceptibility of the off-stoichiometric samples with Fe excess ($x>0$) exhibits a significant field dependence at high temperatures, indicating additional magnetic contribution due to the second phase revealed by x-ray diffraction. It manifests pronounced magnetic irreversibility below 80 K with a clear splitting of ZFC and FC curves and the appearance of magnetic hysteretic effects. The second phase shows a transition into the paramagnetic state at temperatures above 580 K.

4. The specific heat of the stoichiometric samples ($x=0$) and the off-stoichiometric samples ($x=0.06$) is dominated by the lattice contribution at temperatures above 50 K. Below 50 K, down to the lowest temperatures no sharp anomaly in the specific heat characteristic of long-range magnetic or structural transformations is detected.

5. Below 20 K, both stoichiometric and off-stoichiometric samples develop a similar broad feature in the specific heat which can be reasonably well explained by a contribution from the low-lying electronic levels of $Fe^{2+}$ ions with entangled spin and orbital degrees of freedom of the main $FeSc_2S_4$ phase.

6. The obtained results provide clear evidence for the strong influence of excess Fe on the intrinsic magnetic behavior of $FeSc_2S_4$ and confirm the spin-orbital liquid scenario proposed in earlier studies for stoichiometric compound.

**Acknowledgements**

The authors thank Dana Vieweg for experimental support and Oksana Zaharko for fruitful discussions. This work was supported by the Transregional Research Collaboration TRR 80 (Augsburg, Munich, and Stuttgart) and SCOPES project IZ73Z0_152734/1 of the Swiss National Science Foundation.

*Supplemental Material*

Here we present: 1) details of the single crystal growth; 2) results of the measurements of the composition of stoichiometric and off-stoichiometric samples; 3) susceptibility and remnant magnetization data for the off-stoichiometric sample demonstrating magnetic irreversibility effects; 4) experimental data and fits to the specific heat for the non-magnetic spinel $ZnSc_2S_4$.

**S1. Single Crystal Growth**

For the growth of single crystals the chemical transport-reactions method was chosen, which is known as a superior method to obtain high-quality, structurally perfect and impurity-free samples of incongruently melting spinel compounds. In other methods, for example, flux growth, it is difficult to escape from the flux inclusions, which can severely modify the sample properties and mask the intrinsic behavior. The growth of $FeSc_2S_4$ single crystals by chemical transport was a challenging task. Beside the problems of maintaining stoichiometry it was necessary to overcome technical difficulties related to strong interactions of the ternary material with the quartz glass ampoule at high temperatures. The best results in growing stoichiometric single crystals of $FeSc_2S_4$ were obtained by using iodine as transport agent. We also tried several other transport agents, e.g., chlorine and bromine. However in those cases the samples were heavily doped by the transport agents. The growth process was performed in double quartz ampoules. The inner ampoule was graphitized to reduce interaction of the material with the walls of the ampoule. The outer ampoule was filled with pure argon gas at a pressure of 0.9 bar. The growth experiments were performed in a two-zone furnace at temperatures between 900 and 950 °C for 8 to 12 weeks. At the end of the growth cycles, the ampoules were directly removed from the hot furnace by quenching in air, to prevent oxidation of the crystals due to explosion of the ampoules on cooling when using a conventional switch-off regime of the furnace.

In Fig. S1 (a-c) the images of the grown single crystals are shown for several growth runs. The crystals had octahedron- and plate-like shapes with shiny surfaces and dimensions between 1 and 3 mm. The irregular shape of some crystals results from the brake of the ampoule on cooling.

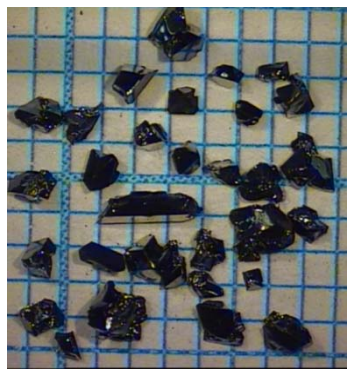 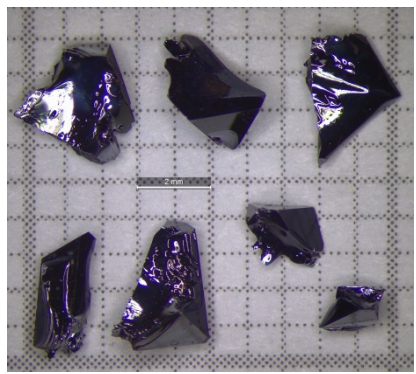 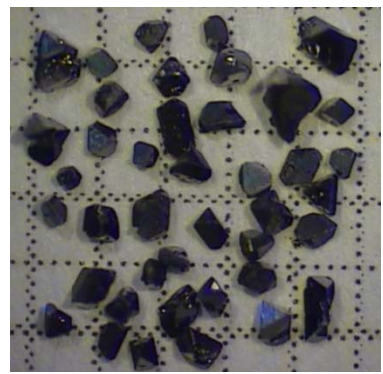

Fig. S1 (a) Image of single crystals grown in the run ATR 236. Grid 1mm.

Fig. S1 (b) Image of single crystals grown in the run ATR 274. Grid 1mm.

Fig. S1 (c) Image of single crystals grown in the run ATR 268. Grid 1mm.

**S2. Composition analysis by WDS**

The chemical composition of the samples was determined with an electron-probe microanalyzer CAMECA SX 50 using wavelength dispersive x-ray spectroscopy (WDS). The measurements were performed on plane parallel single crystalline samples with carefully polished surfaces. The measurements on the stoichiometric polycrystalline sample were done on dense pressed pellets with polished surfaces. High-purity elemental Fe (99.99%) was used as a standard of iron. As a standard of S, stoichiometric CdS, ZnS and $FeCr_2S_4$ single crystals were used. As a standard of Sc, single crystalline $FeSc_2S_4$ samples with well-defined stoichiometry were used. In these samples the concentrations of Fe and S were measured first and the Sc concentration was calculated from the difference. The measurements of these samples were repeated several times until statistical deviations for Sc better than 2% were obtained. The estimated errors of determination of the absolute element concentrations were not higher than 1.5% for Fe and 2% for Sc and S. For single crystalline samples the data were collected under magnification 1600 and averaged over multiple measured spots (minimum 10) with an area of $80 \times 60$ $\mu m^2$. For the polycrystalline sample the data were averaged over the points measured (under magnification 16000) on 15 to 20 different single-crystalline grains of about 40 μm in diameter. Representative data for the polycrystalline sample with $x=0$ and one of the single crystalline samples with Fe excess are given in Table 1SM and 2SM, respectively

.

Table 1SM. Data of the WDS analysis for $FeSc_2S_4$ polycrystal ATF 526 ($x=0$).

| Point | S (W%) | Sc (W%) | Fe (W%) | Sum (W%) | S (A%) | Sc (A%) | Fe (A%) |
|---|---|---|---|---|---|---|---|
| 1 | 47.08 | 33.03 | 20.07 | 100.18 | 57.30 | 28.68 | 14.03 |
| 2 | 46.89 | 33.33 | 21.29 | 101.51 | 56.57 | 28.68 | 14.75 |
| 3 | 45.97 | 34.44 | 19.77 | 100.18 | 56.14 | 29.99 | 13.86 |
| 4 | 47.45 | 33.29 | 20.93 | 101.67 | 57.02 | 28.53 | 14.44 |
| 5 | 46.95 | 33.53 | 20.47 | 100.95 | 56.83 | 28.94 | 14.23 |
| 6 | 47.01 | 33.06 | 20.18 | 100.25 | 57.20 | 28.70 | 14.10 |
| 7 | 46.80 | 33.36 | 20.91 | 101.07 | 56.66 | 28.81 | 14.54 |
| 8 | 47.56 | 33.48 | 21.22 | 102.27 | 56.87 | 28.56 | 14.57 |
| 9 | 47.02 | 32.82 | 20.87 | 100.70 | 57.06 | 28.40 | 14.54 |
| 10 | 46.77 | 32.77 | 20.87 | 100.41 | 56.95 | 28.46 | 14.59 |
| 11 | 46.57 | 33.09 | 20.42 | 100.08 | 56.86 | 28.82 | 14.31 |
| 12 | 45.92 | 34.16 | 19.86 | 99.94 | 56.22 | 29.82 | 13.96 |
| 13 | 46.50 | 33.15 | 20.77 | 100.42 | 56.66 | 28.81 | 14.53 |
| 14 | 47.41 | 32.10 | 20.98 | 100.49 | 57.57 | 27.80 | 14.63 |
| 15 | 47.64 | 33.48 | 21.05 | 102.18 | 56.98 | 28.56 | 14.46 |
| Average | **46.90** | **33.27** | **20.65** | **100.73** | **56.91** | **28.62** | **14.39** |
| Std Dev | 0.50 | 0.47 | 0.55 | 0.83 | 0.41 | 0.48 | 0.27 |

When normalizing the average data on Sc to 2 per formula unit one obtains the composition $Fe_{1.006(19)}Sc_{2.000(33)}S_{3.977(29)}$ corresponding to ideal stoichiometric ratio 1:2:4 of the elements within the errors of their determination.

Table 2SM. Data of the WDS analysis for single crystalline sample ATR 268 ($x$=0.23).

| Point | S (W%) | Sc (W%) | Fe (W%) | Sum (W%) | S (A%) | Sc (A%) | Fe (A%) |
|---|---|---|---|---|---|---|---|
| 1 | 46.58 | 28.97 | 24.91 | 100.46 | 57.12 | 25.34 | 17.54 |
| 2 | 46.70 | 28.85 | 24.57 | 100.12 | 57.38 | 25.28 | 17.34 |
| 3 | 46.38 | 29.01 | 24.70 | 100.09 | 57.08 | 25.46 | 17.45 |
| 4 | 46.45 | 28.67 | 25.13 | 100.25 | 57.11 | 25.14 | 17.74 |
| 5 | 46.34 | 28.84 | 24.50 | 99.68 | 57.23 | 25.40 | 17.37 |
| 6 | 46.27 | 28.82 | 25.10 | 100.19 | 56.96 | 25.31 | 17.74 |
| 7 | 45.69 | 28.88 | 24.56 | 99.13 | 56.84 | 25.62 | 17.54 |
| 8 | 46.40 | 29.01 | 24.44 | 99.85 | 57.20 | 25.50 | 17.30 |
| 9 | 46.18 | 29.22 | 25.21 | 100.60 | 56.67 | 25.57 | 17.76 |
| 10 | 46.26 | 29.17 | 25.14 | 100.57 | 56.76 | 25.53 | 17.71 |
| **Average** | **46.40** | **28.94** | **24.83** | **100.20** | **57.04** | **25.42** | **17.55** |
| Std Dev. | 0.16 | 0.17 | 0.30 | 0.31 | 0.23 | 0.15 | 0.18 |

The average data for this sample correspond to composition $Fe_{1.225(13)}Sc_{1.775(11)}S_{3.982(16)}$ which can be presented as $0.882FeSc_2S_4+0.362Fe_{0.9}S$ for the case of two coexisting phases in the sample.

### S3. Magnetic properties

Figures 2SM show the temperature dependences of the magnetic susceptibility (a) and of the inverse susceptibility (b) for polycrystalline sample with iron excess $x$=0.24 measured in different applied magnetic fields. A strong field dependence of the susceptibility is a generic feature of samples with iron excess. In addition, the susceptibility measured in low fields shows a clear splitting between ZFC and FC data with a maximum at $T_m$≈10 K indicating magnetic inhomogeneity, which we attribute to the presence of the second magnetic phase. Note also high non-linearity of the inverse susceptibility measured on low fields and reduction of the susceptibility due to saturation of the ferromagnetic-like magnetization by a field of 5 T.

Fig. 3SM shows the temperature dependences of the remnant magnetization $M_r$ for polycrystalline sample with iron excess $x$=0.24 performed in two different sequences. The measurements in sequence 1 were performed on warming after cooling in zero field to a temperature of 2 K, then applying the maximum field of 5 T, reducing the field to zero and measure remnant magnetization on increasing temperature. In these conditions, $M_r$ exhibits a sharp drop on increasing temperature from 2 to 10 K and a shoulder at around 20 K. The measurements in sequence 2 were done also on warming but after the measurements of the magnetization up to 5 T at 20 K, reducing the field to zero and lowering the temperature to 2 K. Sequence 2 allows to point out the feature in $M_r$ developing at 20 K which is related to the second magnetic phase existing in samples with Fe excess.

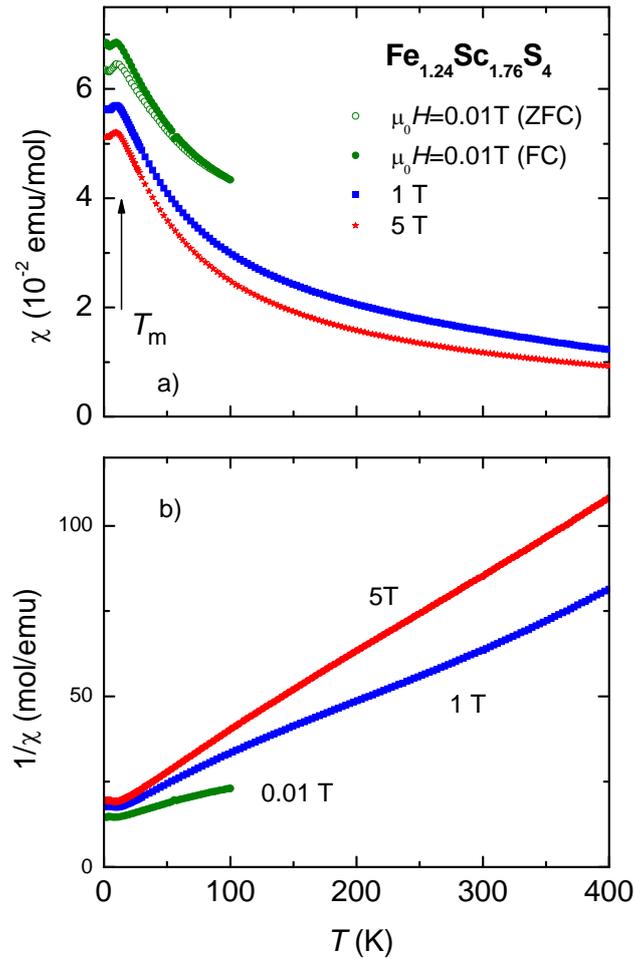

Fig. 2SM (color online). Temperature dependences of the magnetic susceptibilities (a) and of the inverse susceptibilities (b) for polycrystalline sample with iron excess $x=0.24$ measured in different magnetic fields.

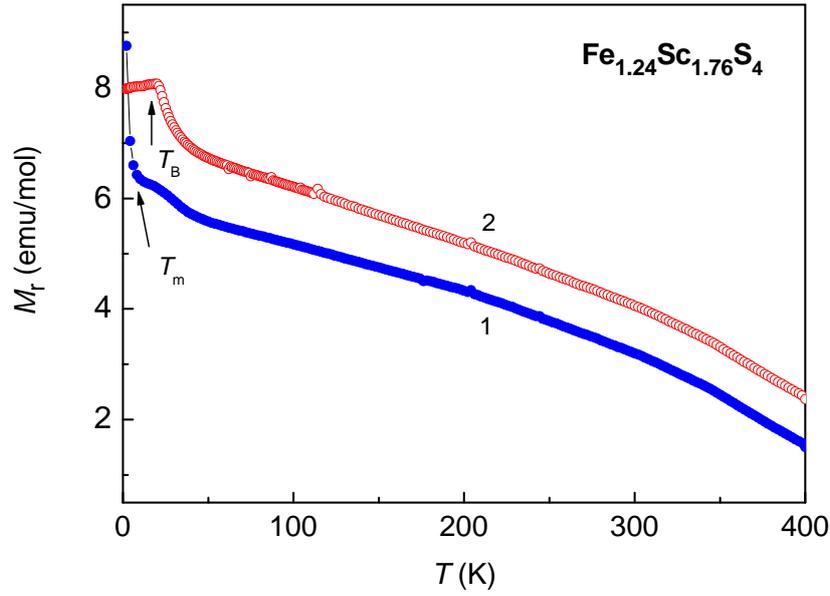

Fig. 3SM (color online). Temperature dependences of the remnant magnetization for polycrystalline sample with iron excess $x=0.24$ measured in different sequences as described in the text.

## S5. Specific heat

In Fig. 4SM the specific heat plotted as $C/T$ vs. $T$ for stoichiometric $FeSc_2S_4$ together with two non-magnetic spinels $ZnSc_2S_4$ and $CdIn_2S_4$ are shown. The advantage of the specific heat data for $ZnSc_2S_4$ for the estimation of the lattice contribution to the specific heat $C_{lat}$ of $FeSc_2S_4$ is evident.

We fitted the specific heat of $ZnSc_2S_4$ by one Debye ($D$) and three Einstein terms ($E1 - E3$). The 21 degrees of freedom per formula unit were taken into account by setting the ratio of these contributions to $D : E1 : E2 : E3 = 1 : 1 : 3 : 2$ using the Debye and Einstein temperatures $\theta_D = 179$ K, $\theta_{E1} = 173$ K, $\theta_{E2} = 334$ K, and $\theta_{E3} = 487$ K. Solid line in Fig. 4SM presents the fit to the experimental specific heat of $ZnSc_2S_4$. Using this combined Debye-Einstein model we were able to describe the specific heat of $ZnSc_2S_4$ within the accuracy better than 0.5% in the temperature range from 40 to 250 K. By introducing corrections to the Debye and Einstein temperatures for $FeSc_2S_4$ taking into account the difference in the molar mass compared to $ZnSc_2S_4$ (see main text) we arrived at a total entropy of $R\ln10$ for the magnetic contribution to the specific heat of $FeSc_2S_4$. Note that the estimations of the lattice contribution based on the specific heat data for heavier $CdIn_2S_4$ performed in Refs. 15 and 16 resulted in a significantly lower value ($\sim R\ln4$) of the entropy contained in the magnetic specific heat of $FeSc_2S_4$ due to overestimation of $C_{lat}$.

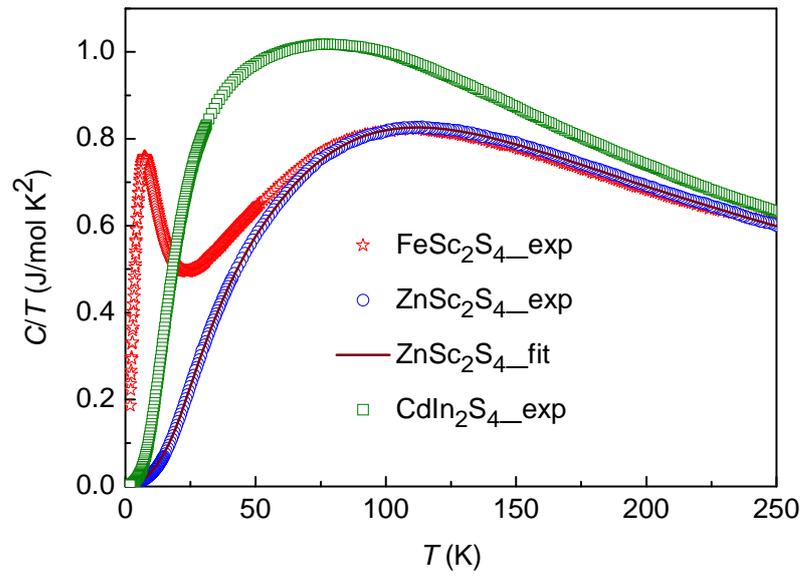

Fig. 4SM (color online). Specific heat plotted as *C/T vs. T* for several measured spinel samples: stoichiometric FeSc$_2$S$_4$, non-magnetic ZnSc$_2$S$_4$ and CdIn$_2$S$_4$. Solid line marks the fit to the experimental specific heat of ZnSc$_2$S$_4$.